\documentclass{aa}
\usepackage{txfonts}
\usepackage{graphicx}
\usepackage{amssymb}
\usepackage{natbib}
\begin{document}
%
%\title{Distance to Atlas, a double lined spectroscopic and
%              interferometric double star in the Pleiades}
\title{A purely geometric distance to the binary star Atlas, a member
of the Pleiades\thanks{Based on observations made with the ELODIE echelle
spectrograph mounted on the 1.93-m telescope at the Observatoire de
Haute-Provence (CNRS), with the FEROS echelle spectrograph mounted on the 2.2-m
telescope at ESO-La Silla Observatory (programme No 072.D-0235B), with the
CORALIE echelle spectrographe mounted on the 1.2-m Euler Swiss telescope at
ESO-La Silla Observatory, with the Naval Prototype Optical Interferometer (US
Naval Observatory) and with the Mark\,\textsc{iii} stellar interferometer at
Mt Wilson.}}
\author{N.~Zwahlen\inst{1} \and P.~North\inst{1}
\and Y.~Debernardi\inst{1} \and L.~Eyer\inst{2} \and F.~Galland\inst{3}
\and M. A. T.~Groenewegen\inst{4} \and C. A.~Hummel\inst{5}}
\offprints{N. Zwahlen} \mail{pierre.north@epfl.ch}
\institute{Laboratoire d'Astrophysique de l'EPFL, Observatoire,
           CH-1290 Chavannes-des-Bois, Switzerland \and
           Observatoire de Gen\`eve, CH-1290 Sauverny,
           Switzerland \and
           Laboratoire d'Astrophysique, Observatoire de Grenoble,
	   F-38041 Grenoble Cedex 9, France \and
	   Instituut voor Sterrenkunde, K.U. Leuven, Celestijnenlaan 200B,
	   B-3001 Leuven, Belgium \and
           European Southern Observatory, Casilla 19001, Santiago 19, Chile}
\date{Received xxxxxx/ Accepted xxxxxx}
\authorrunning{Zwahlen et al.}
\titlerunning{Distance to Atlas, a double star in the Pleiades}
\abstract{We present radial velocity and new interferometric measurements of
the double star Atlas, which permit, with the addition of published
interferometric data, to precisely derive the orbital
  parameters of the binary system and the masses of the components.
  The derived semi-major
  axis, compared with its measured angular size, allows to  determine a
  distance to Atlas of $132\pm 4$~pc in a purely geometrical way.
  Under the assumption that the location of Atlas is representative of the
  average distance of the cluster, we confirm the distance value generally
  obtained through main sequence fitting, in
  contradiction with the early Hipparcos result (118.3$\pm 3.5$~pc).
   }

\maketitle

\def\cms{\,cm\,s$^{-1}$}      %cm s-1
\def\ms{\,m\,s$^{-1}$}        %ms -1
\def\kms{\,km\,s$^{-1}$}      %kms -1
\def\vsini{$v$\,sin\,$i$}     %vsini
\def\m2s2{\,m$^{2}$\,s$^{-2}$} %m2 s-2

\section{Introduction}
The distance of the Pleiades has been under debate ever since the
publication of Hipparcos parallaxes (ESA 1997). Hipparcos data
allowed to precisely calibrate and compare the positions of open
cluster main sequences with those obtained using the usual main sequence
fitting technique, and resulted in a mismatch in the case of at least two
clusters (Pleiades and Coma Ber), the most
severe being that of the Pleiades \citep{MP98}.
Hipparcos locates the Pleiades at $118.3\pm 3.5$ ~pc \citep{FvL99}, about 10\%
nearer than usually quoted results \citep[$132\pm 4$~pc, see e.g.][]{GM93},
that is about 14 pc nearer, or $\sim 0.3$~mag brighter in distance modulus,
or $\sim 1$~mas larger in parallax.

Various explanations have been proposed to solve the problem:
large helium abundance \citep{M97}, metal deficiency \citep{MG00}, problem in
Hipparcos parallaxes \citep{M02}. None of them brought scientific
consensus.

The elements in the balance are not minor ones, as pointed out by
\citet{BP03}. On one side, stellar evolution theory and photometric
calibrations are at stake, or, on the other side, problems appear when
averaging parallaxes in order to reach sub milliarcsecond accuracy from
Hipparcos data. The problem must be solved, especially in the perspective of
the future Gaia mission, which will acquire astrometric data with the same
principle as Hipparcos. Thus, identifying the source of the problem is
worth the effort. It is perhaps relevant to note here that the Pleiades lie
near the ecliptic, where the number of great circles abscissae used for
astrometry is especially small.

Fortunately, there exists several other ways to determine distances
with no or little model dependencies.

This paper presents radial velocity (hereafter RV) measurements of the
interferometric
binary Atlas, which permit to derive the distance of Atlas from purely
geometric arguments. This star was first discovered to be a binary
%(with a magnitude difference of $1.63$ in the $B$ filter)
by \citet{PB75} from
occultation by the Moon. More recent interferometric measurements by
\citet{XP04} yielded well constrained orbital parameters
%: the period was found
%to be $290.81\pm 0.06$~days, the semi-major axis $12.94\pm 0.11$~mas, and
%the eccentricity $0.2457\pm 0.006$. \citet{XP04} derived a distance of
%$135\pm 2$~pc from these parameters and from a mass-luminosity relation,
which lead to a distance of $135\pm 2$~pc, but only when combined with a
mass-luminosity relation, making the result model dependant. On the contrary,
our RV data makes our work completely model free.

Another star allowed an independent measurement of the distance to the
Pleiades. Using Hipparcos epoch photometry, \citet{GT03} discovered that
the spectroscopic binary HD~23642 (HIP~17704) is also eclipsing.
\citet{UM04} obtained photometric and additional RV observations
and derived a distance of $132\pm 2$~pc, perfectly compatible with the result
of the main sequence fitting technique, even though the error bar looks
rather optimistic. This method is almost model-free, but not entirely
since the effective temperatures of the components were determined through
comparison with synthetic spectra.

In addition, Soderblom et al. recently announced (AAS meeting) a distance
of $135\pm 3$~pc on the basis of new parallax measurements
made with the HST on three stars.

\section{Observations and data reduction}
\subsection{Additional interferometric observations:}
\begin{figure}
 \resizebox{\hsize}{!}{\includegraphics{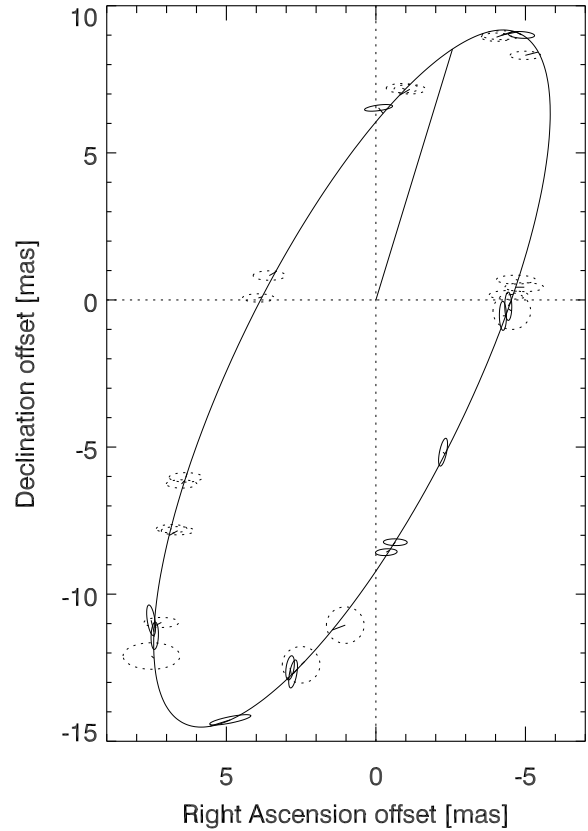}}
 \caption{Astrometric orbit with published data from \citet{XP04} (dotted
 error ellipses) and new data from C. Hummel (full lines;
 see Table~\ref{jastro}). The ellipses indicate the errors on
 each axis, while the full straight line denotes the periastron.}
 \label{orbastro}
\end{figure}

Atlas was observed by CH between 1989 and 1992 with the Mark\,\textsc{iii}
interferometer,
and in 1997 and 2000 with the Navy Prototype Optical Interferometer (NPOI).
The journal of these astrometric observations is given in Table~\ref{jastro}.
The reduction and calibration of the data followed the procedures
described by \citet{H98}.
Although these data are less numerous than those published by \citet{XP04}
on the basis of observations with the Mark\,\textsc{iii} stellar interferometer and
with the Palomar Testbed Interferometer (PTI), those collected with the NPOI
have the advantage of being free from any ambiguity on the orientation of the
orbit, because NPOI measures closure phases. The precision is comparable with
that reported by \citet{XP04}, so we could merge both samples together and
obtain an improved estimate of orbital parameters (see Table~\ref{orbit} below),
using the OYSTER code. The observations
and fitted orbit are shown in Fig.~\ref{orbastro}.
\begin{table}
\caption{New astrometric measurements taken at the Mark\,\textsc{iii}
(BJD $< 2450000$) and NPOI (BJD $> 2450000$) interferometers.
BJD means Barycentric Julian Day. Column 2 gives the number of visibilities,
while the last three columns give the axes and orientation of the error
ellipse.}
\begin{center}
\begin{tabular}{ccrrrrr}
\hline
BJD &Nb. of&$\rho$ & $\theta$ &$\sigma_{\rm maj}$&$\sigma_{\rm min}$&
$\varphi$ \\
$-2400000$&visib.&(mas)&($\deg$)&\multicolumn{2}{c}{(mas)}&($\deg$)\\ \hline
47833.8408& 12& 15.08& 161.16&  0.70&  0.11&101.6\\
48532.8198& 58&  6.53& 359.22&  0.47&  0.10& 97.7\\
48561.8206& 74& 10.24& 331.61&  0.44&  0.11& 86.6\\
48940.8405& 58&  8.26& 184.50&  0.40&  0.11& 88.9\\
48941.8267& 82&  8.58& 182.32&  0.36&  0.11& 93.0\\
50787.8002&418& 13.60& 147.06&  0.47&  0.12&178.0\\
50788.7864&240& 13.23& 145.38&  0.53&  0.13&  8.5\\
51813.7875&644&  4.43& 267.23&  0.48&  0.12&  0.3\\
51815.7963&644&  4.28& 262.75&  0.49&  0.12&180.6\\
51835.7755&368&  5.64& 203.45&  0.48&  0.12&168.5\\
51877.7793&184& 13.01& 167.71&  0.48&  0.12&169.8\\
51879.7881&184& 12.83& 167.07&  0.41&  0.13&171.7\\ \hline
\end{tabular}
\end{center}
\label{jastro}
\end{table}
\subsection{Spectroscopic observations:}
Echelle spectra of Atlas were obtained using above all the ELODIE spectrograph
attached to the 1.93m telescope at Observatoire de Haute-Provence, France,
but also with the CORALIE spectrograph attached to the Swiss Euler 1.2m
telescope at ESO La Silla, Chile, and with the FEROS spectrograph attached
to the 2.2m telescope at ESO La Silla. Almost all observations were gathered
during the 2003-2004 season, though a few archive spectra taken with ELODIE
date as far back as 1995. All spectra were taken without simultaneous ThAr
calibration, either in star mode only, or in star+sky mode. Until the end of
February 2004, observations were made within time normally allocated to other
programmes, thanks to the good will of some observers. From the 27th of
February to the 29th of March 2004, spectra were taken at OHP within officially
allocated time in service mode. The exposure times were typically $10-15$~mn,
resulting in a S/N ratio varying from $\sim 120$ in the worst atmospheric
conditions to more than 400 in the best ones.

The ELODIE and CORALIE spectra were reduced and normalized to continuum
using IRAF, with the method described by \citet{EN02}. The FEROS spectra were
reduced using the standard reduction pipeline provided by ESO in the MIDAS
context. Then, a more careful continuum normalization was performed in the
immediate vicinity of the few lines which appeared useful for RV
determination.

The journal of the observations is given in Table~\ref{jspectro}. Since
spectra with $S/N < \sim 200$ proved to be of little help because of the
shallowness of the lines ($< 0.1$), only the best exposed ones are
listed here.
\begin{table}
\caption{Journal of the specroscopic observations. Epochs are given in 
Barycentric Julian Day. Key to the observers: yd=Y. Debernardi,
cv=C. Vuissoz, fg=F. Galland, mg=M. Groenewegen, rm=R. Monier, pn= P. North,
rds=R. da Silva, bp=B. Pernier, pg=P. Girard}
\begin{center}
\begin{tabular}{ccccrc}
\hline
Date &BJD & Instr. & expos.  & S/N & obs. \\
     &$-2400000$&  & time (s)&     &      \\ \hline
1995.11.08 & 50030.4725 & ELODIE  &  361 & 342 & ? \\
2001.08.04 & 52126.6265 & ELODIE  &  268 & 198 & yd \\
2003.09.14 & 52897.6390 & ELODIE  &  450 & 419 & cv \\ 
2003.11.05 & 52949.5254 & ELODIE  &  600 & 294 & fg \\
2003.11.14 & 52958.4601 & ELODIE  &  600 & 291 & bp \\
2003.11.17 & 52961.4405 & ELODIE  &  600 & 381 & bp \\
2003.12.11 & 52985.6503 & CORALIE &  900 & 223 & mg \\
2004.01.05 & 53010.2836 & ELODIE  &  600 & 351 & rm \\
2004.01.10 & 53015.3717 & ELODIE  &  600 & 324 & rm \\
2004.02.02 & 53038.3266 & ELODIE  &  300 & 314 & fg \\
2004.02.27 & 53063.2736 & ELODIE  &  900 & 279 & fg \\
2004.02.29 & 53065.2571 & ELODIE  &  900 & 218 & fg \\
2004.03.02 & 53067.2649 & ELODIE  &  600 & 316 & rds \\
2004.03.03 & 53068.2741 & ELODIE  &  600 & 303 & rds \\
2004.03.05 & 53070.4947 & FEROS   &  250 & 315 & pn \\
2004.03.06 & 53071.4943 & FEROS   &  250 & 313 & pn \\
2004.03.07 & 53072.2627 & ELODIE  &  600 & 246 & rds \\
2004.03.07 & 53072.4966 & FEROS   &  250 & 312 & pn \\
2004.03.10 & 53075.2677 & ELODIE  &  600 & 296 & pg \\
2004.03.11 & 53076.2775 & ELODIE  &  600 & 332 & pg \\
2004.03.14 & 53079.2716 & ELODIE  &  900 & 440 & pg \\
2004.03.15 & 53080.2739 & ELODIE  &  900 & 460 & pg \\
2004.03.27 & 53092.3056 & ELODIE  &  900 & 270 & fg \\
2004.03.28 & 53093.3090 & ELODIE  &  900 & 340 & fg \\ \hline
\end{tabular}
\end{center}
\label{jspectro}
\end{table}
\section{Analysis of spectra and orbit determination}
Since the lines of the components are wide and always blended, we have used
the KOREL code by \citet{H95} for spectral disentangling and RV
determination. Even with this tool, the task was a delicate one because of the
very small number of suitable lines and because of their shallowness.
Unfortunately, the Balmer lines could not be used because of their larger width
compared with the relatively small RV amplitude.
\begin{table}
\caption{Lines used in this work for radial velocity estimations.}
\begin{center}
\begin{tabular}{crl}
\hline
Species &$\lambda_0$ [\AA] & Remark \\ \hline
Mg\,\textsc{ii}&4481.2     &	    \\
Si\,\textsc{ii}&6347.1     & telluric lines present\\
Si\,\textsc{ii}&6371.4     & telluric lines present\\ \hline
\end{tabular}
\end{center}
\label{lines}
\end{table}
The lines we used are listed
in Table~\ref{lines}. The cleanest is that of Mg\,\textsc{ii}. The two
Si\,\textsc{ii} lines are appropriate as well, but
suffer from contamination by telluric lines which, although not very deep, do
hamper convergence if they are not corrected for. In order to suppress these
telluric lines, we used the high resolution spectrum of telluric lines provided by
\cite{HWVH00} together with their visible atlas of the Arcturus spectrum. This
spectrum was convolved with a gaussian instrumental profile corresponding to
the resolving power of ELODIE $R=42000$; since CORALIE and FEROS have only
slightly better resolving powers ($R=50000$ and $48000$ respectively), the same
convolved telluric spectrum proved useful for them as well. Since airmass and
humidity varies from one spectrum to the next, the correction had to be
adjusted visually for each spectrum by scaling the telluric line intensities
\citep[see][ for details]{NBL94}, mainly on lines lying on the continuum.
This procedure proved quite satisfactory,
thanks to the fact that the relative depths of the telluric lines remain the
same in a given spectrum even though their absolute depths change from one
spectrum to the other.

Fortunately, most orbital parameters are known from interferometry, especially
$T_0$, $e$, $i$ and $\omega$, so these quantities could be fixed while running
KOREL, leaving only the RV amplitude $K_1$ and the mass ratio $q$ as
adjustable parameters.

\begin{figure}
 \resizebox{\hsize}{!}{\includegraphics{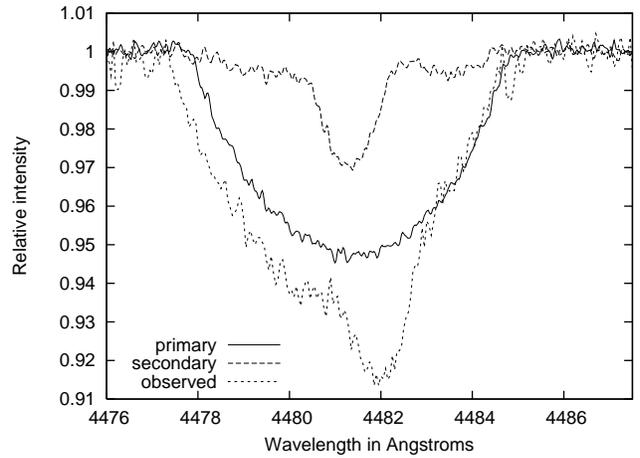}}
 \caption{Disentangled lines of Mg\,\textsc{ii} centered on the systemic
 velocity. Full line: primary; broken line: secondary. Dotted line: spectrum
 observed in 1995 near quadrature, at phase $\phi\sim 0.1$.}
 \label{dispec}
\end{figure}

\begin{figure}
 \resizebox{\hsize}{!}{\includegraphics{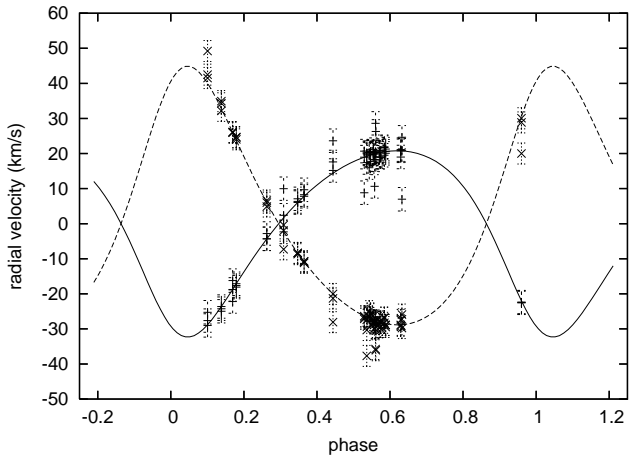}}
 \caption{Radial velocity curves given by the KOREL code. All orbital parameters
 were fixed at their astrometric values, except $K_1$ and $q$. There are three
 radial velocity estimates per epoch, one per line, even though the RV curves
 were constrained simultaneously by all three lines. Note that the systemic
 velocity is arbitrarily set to zero because the KOREL code cannot determine
 it.}
 \label{rvcurve}
\end{figure}

By limiting the data set to spectra with $S/N\geq 200$, and with starting
values of $K_1$ between $10$ and $60$~\kms and of $q$ between $0.3$ and $1.0$,
we found good convergence towards the values given in Table~\ref{orbit}.
The disentangled spectra are shown in Figure~\ref{dispec} for the
Mg\,\textsc{ii} line. One clearly sees that the primary rotates much faster
than the secondary, the line exhibiting a typical rotational profile. The
$v\sin i$ values are respectively
about $240$ and $60$~\kms. The RV curves computed by the KOREL code
are shown in Figure~\ref{rvcurve}; the points are obtained by cross-correlating
the individual spectra with the disentangled spectra.
The error estimate on the fitted orbital parameters is not straightforward,
since the disentangling method used in the KOREL code does not give it in a
natural way. Therefore, we had recourse to the jackknife method \citep{BE93},
which consists
in removing each spectrum in turn from the whole set of data, and run KOREL
with the remaining 23 spectra. One obtains slightly different values of the
fitted parameters, and their variance multiplied by the number of
spectra gives the final error estimate \citep[e.g.][]{Il03}.

Since KOREL gives individual RV values, we mixed the latter with the
astrometric data of \citet{XP04} and ours to fit simultaneously all orbital
parameters. This is in principle a more elegant method than to fit separately
the spectroscopic data and the astrometric ones. However, in our case one has
to bear in mind that the individual RV values are not mutually independent {\it
stricto sensu}, because they result from a cross-correlation between each
individual spectrum and the components' disentangled spectra. Since the latter
are built on the same data, there is some circularity which tends to reduce the
$O-C$ residuals. In addition, the spectroscopic orbit is not independent from
tbe astrometric one, since the latter was used to constrain most orbital
parameters. Nevertheless, such a consistency check is reassuring in that it
provides the same result as when combining the separate astrometric and
spectroscopic solutions.
\begin{table}
\caption{Astrometric and spectroscopic orbital parameters adjusted
with the OYSTER and KOREL codes respectively. A simultaneous
fit of astrometric and RV data yields essentially the same results. The errors
on the spectroscopic parameters are those obtained through the jackknife method
alone, but the error on $d$ includes the small uncertainty on $a"$.}
\begin{center}
\begin{tabular}{crcr}
\hline
Parameter        & value              &Parameter         & value  \\ \hline
Period [d]       & $290.984\pm 0.079 $& $K_1$ (\kms)     & $26.55\pm 1.41$  \\
$T_0$ [BJD]      & $2450583.0\pm 1.9$ & $K_2$ (\kms)     & $36.89\pm 0.22$ \\
$e$              & $0.2385\pm0.0063$  & $q$              & $0.720\pm 0.036$  \\
$i$ ($\deg$)     & $107.87\pm 0.49$   & $a$ (AU)         & $1.73\pm 0.04$  \\
$\omega$ ($\deg$)& $151.9\pm 2.2$     & $M_1$ ($M_\odot$)& $4.74\pm 0.25$\\
$\Omega$ ($\deg$)& $154.0\pm 0.7$     & $M_2$ ($M_\odot$)& $3.42\pm 0.25$\\
$a"$ (mas)       & $13.08\pm 0.12$    & $d$ (pc)         & $132\pm 4$ \\ \hline
\end{tabular}
\end{center}
\label{orbit}
\end{table}
\section{Distance of Atlas and conclusions}
The orbital parameters obtained above yield a final distance of
$d=132\pm 4$~pc which confirms the ``traditional'' distance obtained through
the main sequence fitting technique.
Thus the result obtained by \citet{XP04} is fully confirmed,
and there is no need for any revision of the stellar evolution models, or
assumed helium content of the Pleiades members, or to postulate a non-solar
metallicity.

Our result has the advantage of relying exclusively on {\it empirical} data,
without the least recourse to any model. The only assumption we made is that of
no intrinsic spectral variability of the components of Atlas, an assumption
which is well justified {\it a posteriori} for the Mg\,\textsc{ii} line at
least, but which will be examined further
in a more complete forthcoming paper.

It is worth mentioning that our result is completely coherent with the
distance of the recently discovered eclipsing
binary HD 23642 \citep{UM04}, within the errors and
within the expected depth of the Pleiades cluster.
Thus, there are now two objects for which a more or less direct distance
estimate puts the Pleiades at about $130-135$~pc, which is clearly different
from the Hipparcos estimate published by \citet{FvL99}. It appears increasingly
urgent to understand what went wrong with that estimate, a difficult task
which is in progress \citep{FvL04}.

\begin{acknowledgements}
  We thank Prof. B.~Paczy\'nski for advice and encouragements, Prof. M.~Grenon,
  Prof. L.~Lindegren and Dr. F.~van Leeuwen for theirs comments.
  We are grateful to Prof. M.~Mayor for agreeing to devote a small part of the
  observing time of his group at OHP to Atlas, and to B.~Pernier
  for performing these observations. We thank also Dr. R.~Monier for two
  observations during his own observing programme. We thank R.~Da Silva and
  P. Girard for service mode observations at OHP in March 2004, and Dr. C. Melo
  for the reduction of a FEROS spectrum. We are grateful to Prof. P. Hadrava
  for his code, to S. Iliji\'c for useful comments and to the referee,
  Dr. M. Pinsonneault, for constructive remarks. The Navy Prototype Optical
  Interferometer is a
collaboration between the US Naval Research Lab and the US Naval Observatory
in association with Lowell Observatory,
and is funded by the Office of Naval Research and the Oceanographer of the
Navy. This study was supported by the Swiss National Science
  Foundation.  
\end{acknowledgements}

\bibliographystyle{aa}
\bibliography{atlas}

\end{document}